\documentclass[aps, pre, preprint]{revtex4-1}
\usepackage{soul} 
\usepackage{hyperref}
\usepackage{amsmath}
\usepackage{amssymb}
\usepackage{graphicx}
\usepackage{psfrag}
\usepackage{booktabs} 
\usepackage{tabularx}
\usepackage{tabu}
\usepackage{longtable}
\usepackage{multirow}
\usepackage{color}
\usepackage[table]{xcolor}
\usepackage{pstool}
\usepackage[absolute,overlay]{textpos}
\usepackage{tcolorbox}
\usepackage[caption=false]{subfig}
\hypersetup{
colorlinks=true,
citecolor=blue,
urlcolor=blue
}

\newcommand{\Tr}{\text{Tr}}
\newcommand{\sinc}{\text{sinc}}

\definecolor{color1}{RGB}{230,230,250}
\definecolor{color2}{RGB}{255,255,224}
\definecolor{green2}{RGB}{0,150,0}

\usepackage[normalem]{ulem}

\begin{document}
\title{Out-of-time-order correlators in bipartite nonintegrable systems}
\author{Ravi Prakash}
\altaffiliation{Current Address: Department of Physics,
Motilal Nehru National Institute of Technology Allahabad, Prayagraj -- 211004, India}
\author{Arul Lakshminarayan}
\affiliation{Indian Institute of Technology Madras, Chennai -- 600036, India}
\begin{abstract}
Out-of-time-order correlators (OTOC) being explored as a measure of quantum chaos, is studied here in a coupled bipartite system. Each of
the subsystems can be chaotic or regular and lead to very different OTOC growths both before and after the scrambling or Ehrenfest time. We present preliminary results on weakly coupled subsystems which have very different Lyapunov exponents. We also review the case when both the subsystems are
strongly chaotic when a random matrix model can be pressed into service to
derive an exponential relaxation to saturation.
\end{abstract}
\maketitle
\section{Introduction}

Much of the work on quantum chaos was traditionally based on the 
 Schr\"odinger equation: stationary state properties, time-evolving states and statistics of spectra. The connection of quantum states or observables with classical trajectories is not straight forward, the correspondence holding till the Ehrenfest time. Initially nearby classical trajectories can deviate exponentially at the rate of the Lyapunov exponent and is one of the important ingredients of  classical chaos. Various signatures or imprints of classical chaos in the stationary properties in the corresponding quantum systems have been known from a long time. Recently, the Heisenberg picture has come into prominence with operator evolution providing the opportunity to connect more intimately to the evolution of classical observables, when a semiclassical limit exists. Operator scrambling and out-of-time-ordered correlators (OTOC) are two quantities that are being currently intensely investigated and can be used to define dynamical features of chaos in the quantum system \cite{Shen-2017, Slagle-2017, Cotler-2017, Campisi-2017, Hashimoto-2017, Fan-2017, Keyserlingk-2018, Nahum-2018, Rozenbaum-2017, Jalabert-2018, Arul-Baker-2018, Rakovszky-2018, Saraceno-2018, Chen-2018, Moudgalya-2018, Haehl-2017, Klaus-2018, Rivas2019,ShashiOTOC2019,Diego2019}.  

OTOC were first studied in the context of semiclassical approximations in the theory of superconductivity \cite{Larkin-1969}. More recently the OTOC has been studied in the context of quantum gravity, AdS-CFT correspondence, field theories and many-body physics, including many-body localization \cite{Fan-2017,  Chen-2017, Slagle-2017, Keyserlingk-2018}. The correlators are useful to quantify quantum chaos by defining a quantum analogue of the Lyapunov exponent, as for chaotic systems the OTOC grow exponentially till the Ehrenfest time with a rate which can be considered as a quantum equivalent of the Lyapunov exponent \cite{Rozenbaum-2017, Maldacena-2016}. This is most apparent on studying the increase of non-commutativity of two (say Hermitian) operators, one evolving  with time. We consider the OTOC to be simply
\begin{align}
\label{eq-otoc}
C(t) = -\frac{1}{2} \left< [A(t),B(0)]^2 \right>,
\end{align}
where $\left< \cdot \right>$ represents the thermal average over an ensemble at temperature $T$. A standard semiclassical argument makes
it plausible that this can increase exponentially with time. As, if $A$ and $B$ are the position and momentum operator the commutator $[x(t), p(0)]^2$ is semiclassically equivalent to the Poisson bracket, $\hbar^2 \{x(t),p(0)\}^2 = \hbar^2(\partial x(t)/\partial x(0))^2$, and this exhibits exponential growth for chaotic systems, i.e, $(\partial x(t)/\partial x(0))^2 \approx \exp(2 \lambda t)$ and reflect the sensitive dependence on initial conditions.

Another interesting fact about OTOC is the conjecture the rate has an upper bound, $\lambda \leq 2\pi k_B T/\hbar$ \cite{Maldacena-2016}. The Sachdev-Ye-Kitaev (SYK) model, a disordered model of Fermions with all-to-all interactions, is one of the maximally chaotic system which saturates the bound \cite{SYK-1993, Kitaev-2015}. Similar bound was found in earlier studies of scrambling of quantum information around a black hole horizon \cite{Shenker-2014, Preskill-2007}.
The exponential growth of $C(t)$ occurs in a time window $t_d<t<t_{EF}$ where $t_d$ is a ``diffusion time scale" before which the growth can be dependent on the operators used and is typically a small time scale that does not scale with the system size, while $t_{EF}$ is 
the Ehrenfest time and could be the time of breakdown of classical-quantum correspondence if a classical limit exists.

While OTOC have been studied extensively for many-body systems 
such as spin chains and coupled harmonic oscillators \cite{ShashiOTOC2019}, it presents intriguing features even for single and few-body systems. Therefore many recent studies have also concentrated on low-dimensional systems with a known semiclassical limit such as the quantum standard map, the quantum bakers map, quantum billiards, the perturbed cat map, the kicked top, which can be viewed as an completely  connected spin model \cite{Rozenbaum-2017, Cotler-2018, Chen-2018, Saraceno-2018, Hamazaki-2018}. All these display the expected exponential growth till the Ehrenfest time,  which scales as $\sim \log N$, where $N$ is the Hilbert space dimension. This follows from the Ehrenfest time scaling as $\log(1/h)/\lambda$, where $h$ is a scaled Planck constant and $\lambda$ is the classical Lyapunov exponent and $N \sim 1/h$.

In the Wigner phase space representation, the commutator is equivalent to the Moyal bracket which is equal to the Poisson bracket in the $\hbar \rightarrow 0$ limit. Beyond the Ehrenfest time, the $\hbar$ corrections start dominating 
and there exists no classical correspondence for the OTOC. This contribution studies a model that is bipartite, the Hilbert space of states has a tensor product structure of two $N$ dimensional
weakly interacting subsystems $\mathcal{H}_1^N \otimes \mathcal{H}_2^N$. It is possible to arrange for various dynamical states of the subsystems, when both are regular, both are chaotic or
when one is regular and one is chaotic. Of these in a sense, the best understood is the case when both subsystems are chaotic and of comparable chaoticity.
Being bipartite it differs fundamentally from single particle models studied so far and provides a bridge to fully many-body systems. Another recent study of OTOC in bipartite systems is in \cite{Rivas2019}. In  particular the operators $A$ and $B$ can initially be local to the two subsystems, that is of the form $O_A \otimes \mathbb{I}_B$ and $\mathbb{I}_A \otimes O_B$, and will therefore commute. At later times, solely due to the interaction and entanglement generated they will not commute and their OTOC will grow. We note that there are several systems that could satisfy 
such requirements, including particles in a quantum dot, spin chains wherein the operators are local over two halves of the 
spins, two large coupled tops or spins. In this scenario, the entanglement created due to the interactions 
drives the OTOC growth and in particular the operator entanglement shows similar behavior. Thus this is the simplest 
multipartite setting in which entanglement is responsible for the OTOC growth and information scrambling.

Spectral transitions and eigenstate entanglement when the subsystems are fully chaotic have recently been studied, these being governed by one dimensionless transition parameter in terms of which the growth of the entanglement is universal \cite{Arul-2001, Arul-2016, Arul-PRL-2016}. We find that 
it is this same transition parameter that is also responsible for the OTOC growth beyond the Ehrenfest time, when both the subsystems are chaotic. Thus the OTOC increases in a universal manner with a time scale which maybe characterized as the scrambling time ($t_s$) at which it saturates. 
We show that OTOC can be modeled with a random matrix ensemble in this time domain. Before the Ehrenfest time, the OTOC increase as $e^{2 \lambda t}$ till a time $\sim \log(1/\hbar)$ and is true only for operators with a  semiclassical limit. In contrast when the subsystems are regular and the interaction is weak, OTOC can increase as a power law in time and as we shall argue $\sim t^2$, for a time that is $\sim 1/\sqrt{\hbar}$. However, this is also strongly operator dependent.

We consider the OTOC in the infinite temperature limit, thus the conjectured quantum Lyapunov exponent has no upper bound. Thus, the OTOC for two operators $A$ and $B$ given in (\ref{eq-otoc}) is $C(t)  =  C_2(t) - C_4(t),$ where
\begin{align}
\label{eq-otoc-2}
C_2(t) & = \text{Tr}\left[A(t)^2 B(0)^2\right], \; \text{and}\;\\
C_4(t) & = \text{Tr}\left[A(t)B(0)A(t)B(0)\right] 
\end{align}
where $C_2(t)$ and $C_4(t)$ are the two- and four-point correlations respectively. Since the process of going forward and backward in time is crucial to diagnose quantum chaos, such as in studies of fidelity \cite{Gorin-2006}, only $C_4(t)$, being an out-of-time ordered correlator, is sufficient to explore chaos in quantum systems. Consider $A$ and $B$ to be Hermitian operators localized to each subsystem respectively, {\it i.e.},
\begin{align}
\label{eq-operator-AB}
A(0) = \mathcal O_1\otimes \mathbb{I},\;
B(0) = \mathbb I \otimes \mathcal O_2,
\end{align}
where $\mathcal O_j$ with $j = 1,2$ are Hermitian operators. The evolution of operator $A$ is given by $A(t) = \mathcal U^{-t} A(0) \mathcal U^t$, for integer
times $t$ and it is typically no longer of a tensor product form and fails to commute with $B(0)$ for $t>0$. Two different OTOC are there for bipartite systems, when the operators are localized in the same or different subsystems. To be explicit, we define
\begin{equation}
C_{AA}(t)=-\text{Tr}[A(t),A(0)]^2, \;\; C_{AB}(t)=-\text{Tr}[A(t),B(0)]^2.
\end{equation}
For weak interactions, the growth of $C_{AA}(t)$ is governed predominantly by the dynamical nature of subsystems but for $C_{AB}(t)$ both local dynamics and interaction play significant role, as it is entirely a result of subsystem entanglement. If $C(t)$ grows exponentially with time in systems with bound spectra this is an indicator of quantum chaos. This
is an equivalent to the classical Lyapunov exponent being positive in a bounded system. It may also be pointed out that while we study this version 
of the OTOC below, another possibility is to take a logarithm before the trace, in other words to average the Lyapunov exponents of individual state expectation values in some complete basis.

\section{OTOC for coupled quantum kicked rotors} \label{sec-qkr}
We wish to study a convenient model for the OTOC of bipartite systems, where each subsystem 
can have a range of dynamical behavior from regular to fully chaotic. Coupled area-preserving maps
present themselves as attractive models to study.
The kicked rotor, or the standard map, is a well-known system that exhibits both integrability and chaos as one changes the kicking strength parameter. We consider two interacting coupled kicked rotors whose classical and quantum dynamics has been studied earlier from points of view of 
Arnold diffusion \cite{Wang-1990, Wood-1990}, interplay of chaos and entanglement \cite{Arul-2001}, higher-dimensional Hamiltonian systems \cite{Richter-2014}, level spacing and entanglement transitions in 
strongly chaotic, weakly interacting systems \cite{Arul-PRL-2016, Arul-2016}. As each of the standard map is studied on the torus phase-space, the quantum dynamics is finite dimensional and the classical dynamics is compact.

The composite form of the Hamiltonian is given by \cite{Arul-PRL-2016}
\begin{align}
H = H_1 \otimes \mathbb I_2 + \mathbb I_1 \otimes H_2 + H_{12}
\end{align}
where $H_j$, with $j = 1,2$, represents Hamiltonian for individual sub-system and $H_{12}$ represents interaction. For kicked rotors, we have
\begin{align}
H_j = \frac{1}{2}p_j^2 +  \frac{1}{4\pi^2} K_j \cos(2\pi q_j)\delta_t, \;\; \text{and},\;\;
H_{12} =  \frac{b}{4\pi^2} \cos(2\pi(q_1 + q_2))\delta_t,
\end{align}
with $\delta_t = \sum_{n = -\infty}^{\infty} \delta(t-n)$, and the parameter $b$ is the interaction. The single rotor is integrable only for vanishing kick strength $K = 0$, and there is  a mixed phase space, with a finite measure of chaotic and stable regions as $K$ increases. Although there is no rigorous proof, it is believed that there is widespread chaos for $K \gg 5$. While the average Lyapunov exponent is monotonically increasing with $K$, small  stable phase space structures can arise for example through homoclinic tangencies. 
The coupled map is considerably harder to visualize, being a four dimensional symplectic map $(q_1,p_1,q_2,p_2) \mapsto (q_1',p_1',q_2',p_2')$.
With periodic (unit period) boundary conditions on all of these variables, the phase-space is a 4-torus, on which the map is ($i=1,2)$:
\begin{subequations}
\label{eq-stdmap}
\begin{align}
q_i^\prime & = q_i + p_i^\prime \;\; (\text{mod}\; 1)\\
p_i^\prime & = p_i + \frac{K_i}{2\pi}\sin(2\pi q_i) + \frac{b}{2\pi}\sin[2\pi(q_1 + q_2)]\;\; (\text{mod}\; 1).
\end{align}
\end{subequations}
Here $q_i$ and $p_i$ are position and momentum coordinates of $i$-th rotor immediately after a kick and $(q_i^\prime, p_i^\prime)$ are the coordinates in phase space immediately after the next kick.


As is well-known, the quantum dynamics of the kicked rotors with  torus boundary conditions occurs in a finite dimensional Hilbert space of dimension say $N$, so that both position and momentum have discrete values. The Hilbert space of two coupled rotors is the tensor product space of dimension $N^2$.
The Floquet operator, $F_{K_j}$ of individual rotors in position basis with $0 \leq n_i \leq N-1$ is
\begin{align}
\label{eq-uone}
\nonumber
& \left<n^\prime \left | F_{K_j} \right | n \right >  = \frac{1}{N}\exp\left( -iN\frac{K_j}{2\pi}\cos\left(\frac{2\pi}{N}(n+\alpha)\right)\right) \\
& \times \sum_{m = 0}^{N-1} \exp\left(-i\frac{\pi}{N}(m+\beta)^2 \right) \exp\left(i \frac{2\pi}{N} (m+\beta)(n-n^\prime)\right),
\end{align}
while the interaction $U_b$ is a diagonal matrix with entries given by
\begin{align}
\label{eq-ub}
\nonumber
\left < n_1 n_2 \left | U_b \right | n_1^\prime n_2^\prime \right > & = \exp\left\{ -i N \frac{b}{2\pi}\cos\left[\frac{2\pi}{N}(n_1+n_2+2\alpha) \right]\right\} \\
& \times \delta_{n_1, n_1^\prime} \delta_{n_2, n_2^\prime}.
\end{align}
The $\alpha$ and $\beta$ are parity and time reversal breaking phases respectively and arise from the boundary conditions for the quantum torus. The parity is preserved for $\alpha = 0$ and broken for other values. Similarly the system is time reversal invariant for $\beta = 0$ and $\beta=1/2$. The time evolution operator for the composite system, quantizing the coupled map in Eq.~(\ref{eq-stdmap}) is then given by,
\begin{align}
\label{eq-floquet}
U = (F_{K_1} \otimes F_{K_2}) U_b.
\end{align}
Since we have position and momentum both to be discrete, it is convenient to have the observables, $\mathcal O_1$ and $\mathcal O_2$ to be constructed from the position and momentum translation operators $T_q$ and $T_p$ defined as $T_{q_i} \left |n_i \right > = \left | n_i+1 \right >$ and $T_{p_i} \left | n_i \right > = \exp\left[2\pi i (n_i+\alpha)/N\right]\left |n_i \right >$. We choose,
\begin{align}
\label{eq-a}
\mathcal O_i = \frac{1}{2} (T_{p_i} + T_{p_i}^\dagger)
\end{align}
so that the classical limit of the observables  $\mathcal O_i$ is simply $\cos(2 \pi q_i)$. We consider the following cases: (a) the uncoupled map is integrable ($K_1 = K_2 = 0$), (b) has a mixed phase-space ($K_1 = 0.5, K_2 = 0.7$), (c) is fully chaotic ($K_1 = 9, K_2 = 10$ and $K_1 = 19, K_2 = 20$), or (d) one is regular and the other subsystem is chaotic ($K_1=0, K_2=10)$. We set $\beta = 0$ and $\alpha = 0.35$ to have time reversal invariance and broken parity in the Floquet operator.

\subsection{Pre-Ehrenfest time regime and Poisson brackets} \label{sec-lyapunov-phase}
\begin{figure}[!h]
\includegraphics[width=0.45\textwidth]{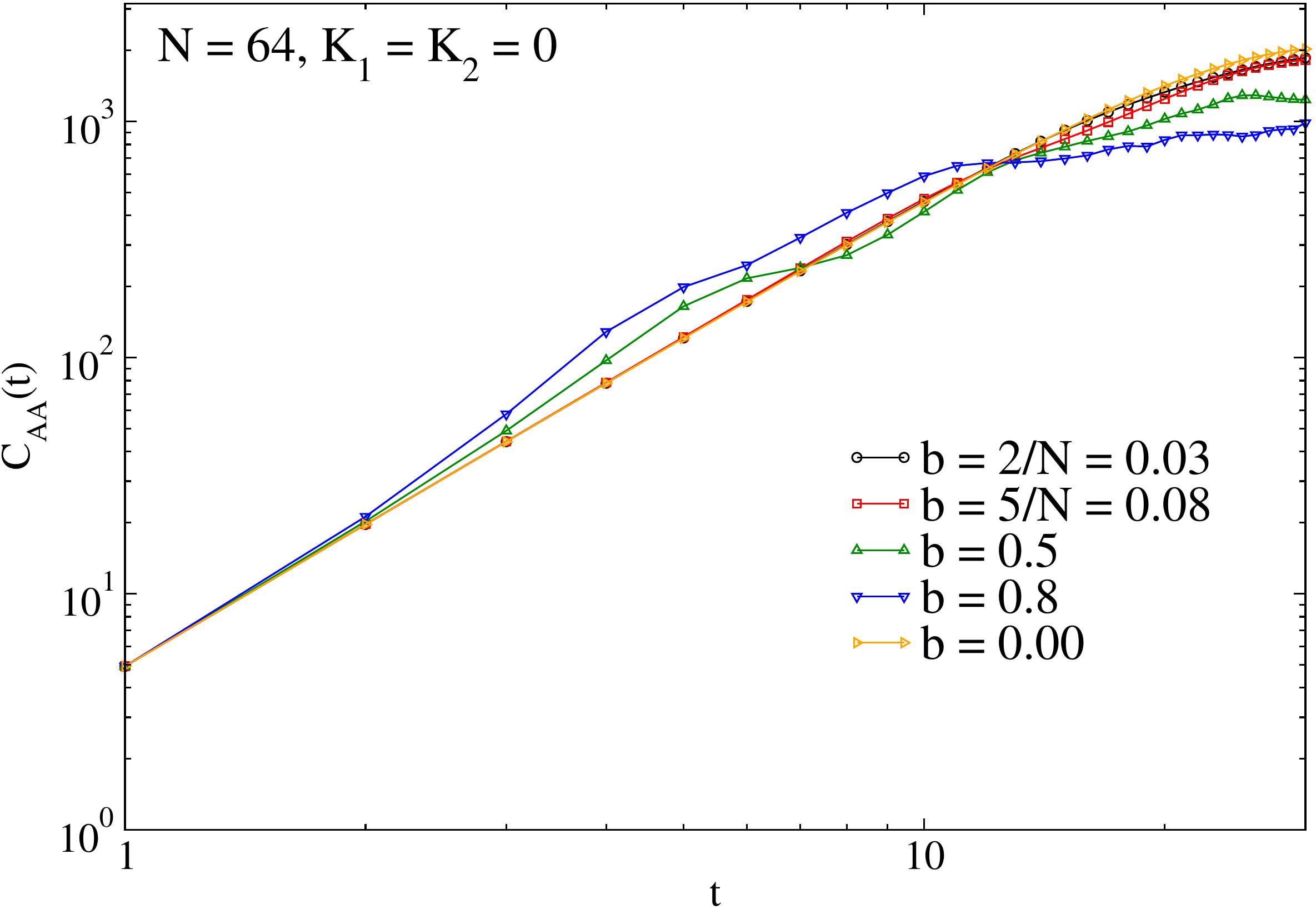} 
\includegraphics[width=0.45\textwidth]{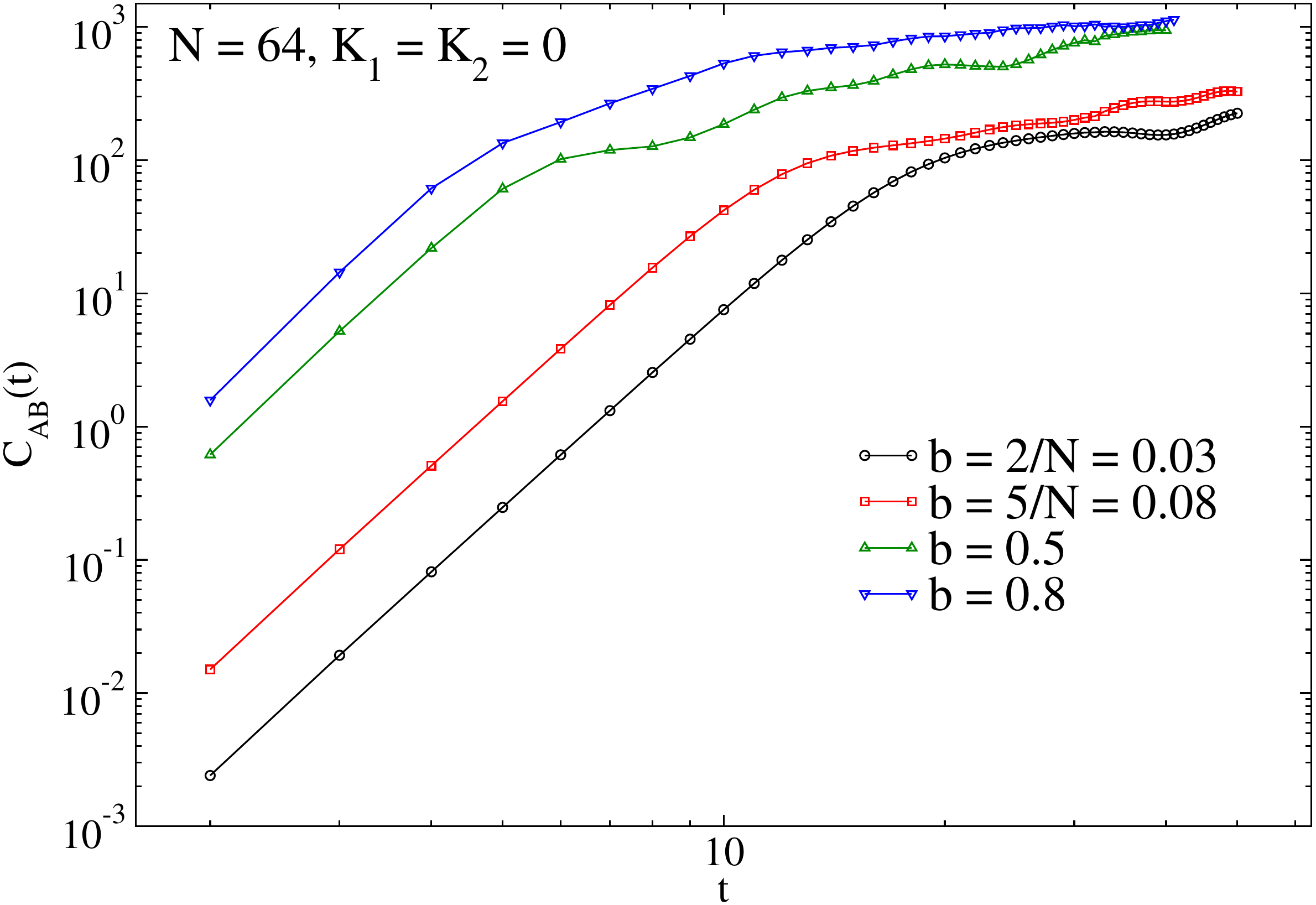}
\caption{The OTOC  when the noninteracting systems are integrable, $K_1=K_2=0$ and the operators are in the same subspace ($C_{AA}$, Left) and in different subspaces ($C_{AB}$, right). Shown are plots in the log-log scale and we see a power law $\sim t^2$ that holds approximately in the first case and a law $\sim t^5$ in the second. }
\label{fig:K00}
\end{figure}

\begin{figure}[!h] 
\includegraphics[width=0.45\textwidth]{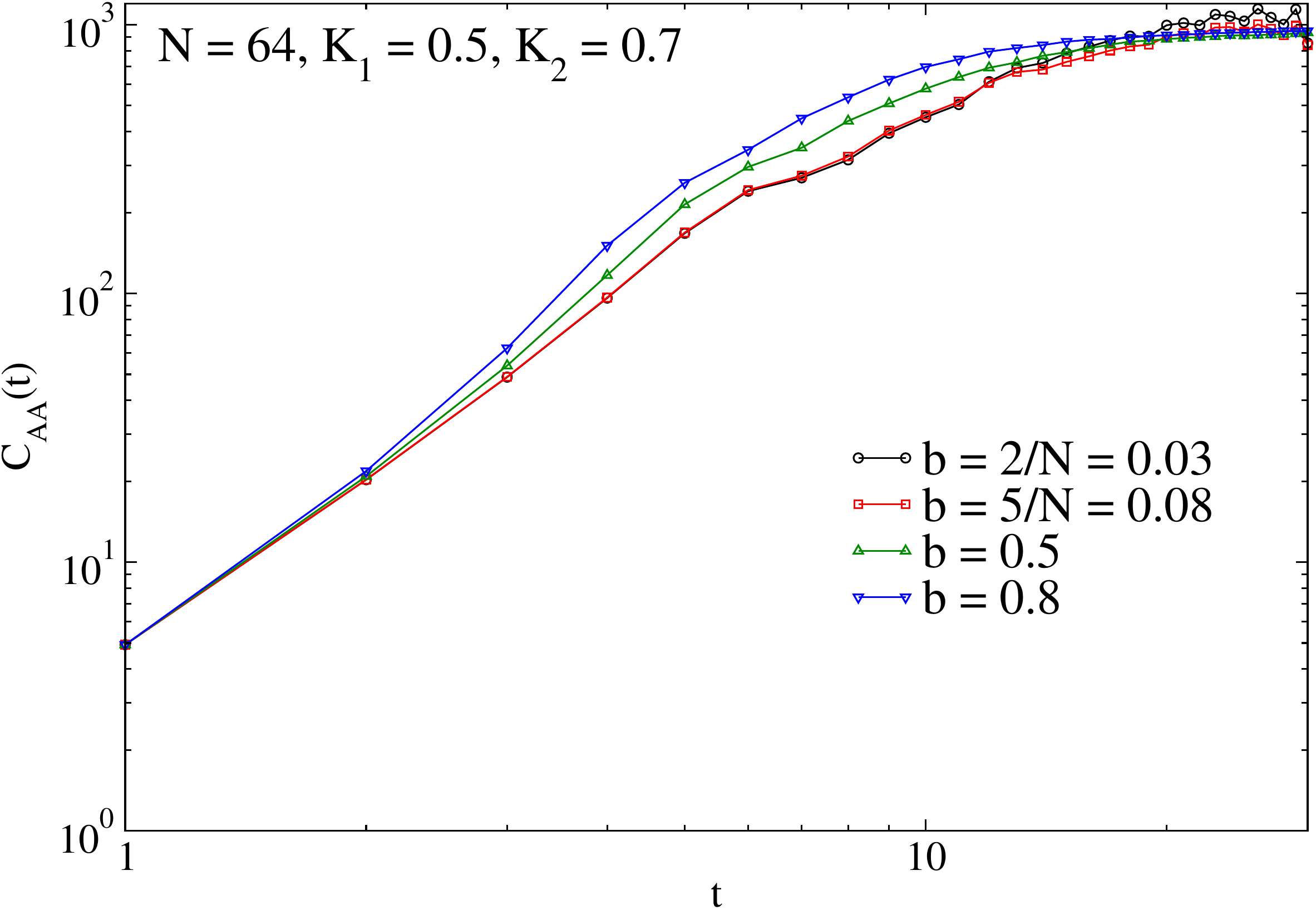}
\includegraphics[width=0.45\textwidth]{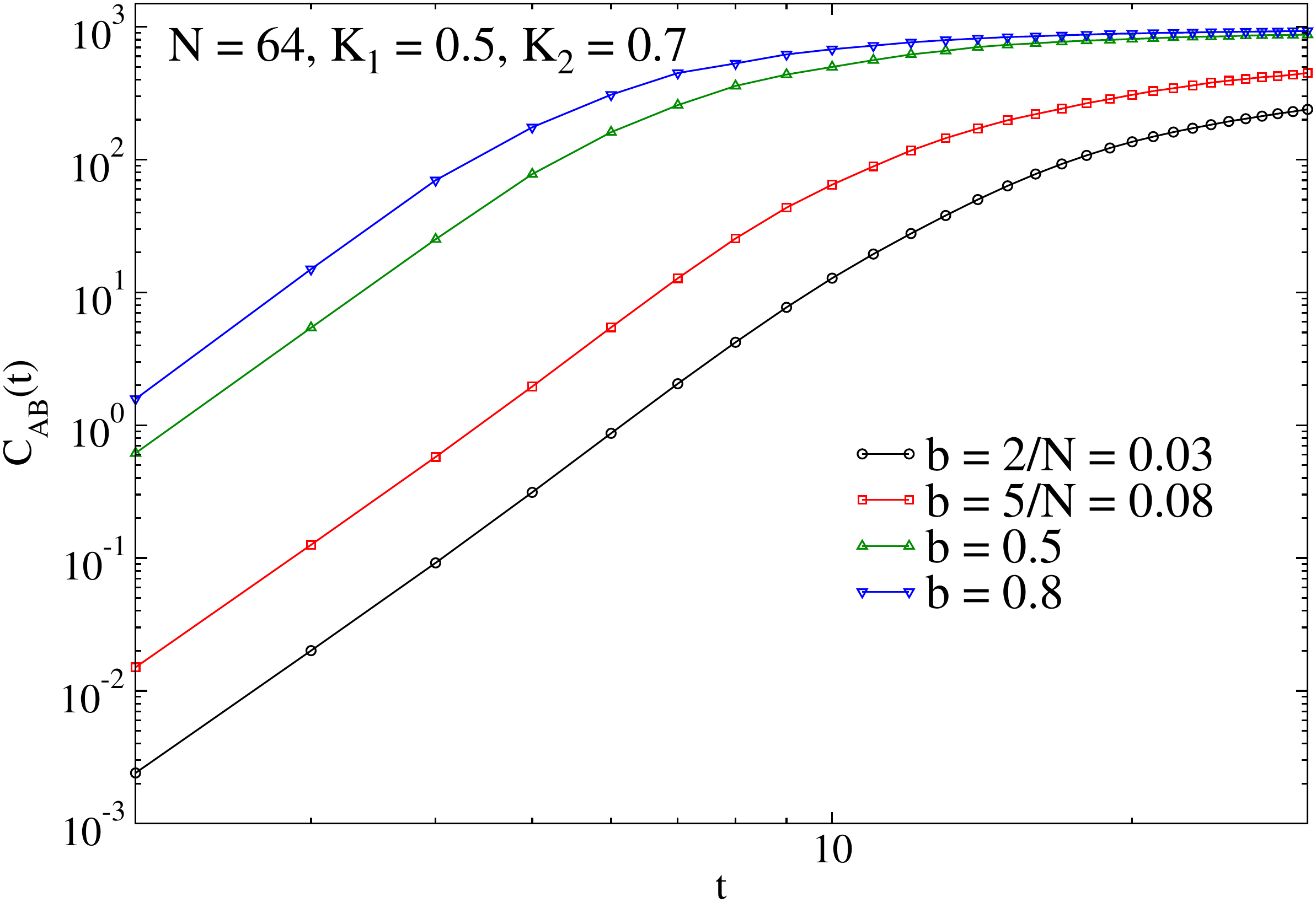}
\caption{$C_{AA}(t)$ and $C_{AB}(t)$ for the weakly chaotic case for several values of interaction strengths. The plots are shown for $K_1 = 0.5, K_2 = 0.7$. The sub-system dimension, $N = 64$ is considered here. The figures indicates that for small $b$ there is a power law growth $\sim t^{5.4}$.\label{fig-otoc-fft-integrable}}
\end{figure}

\begin{figure}[!h]
\centering
\includegraphics[width=0.5\linewidth]{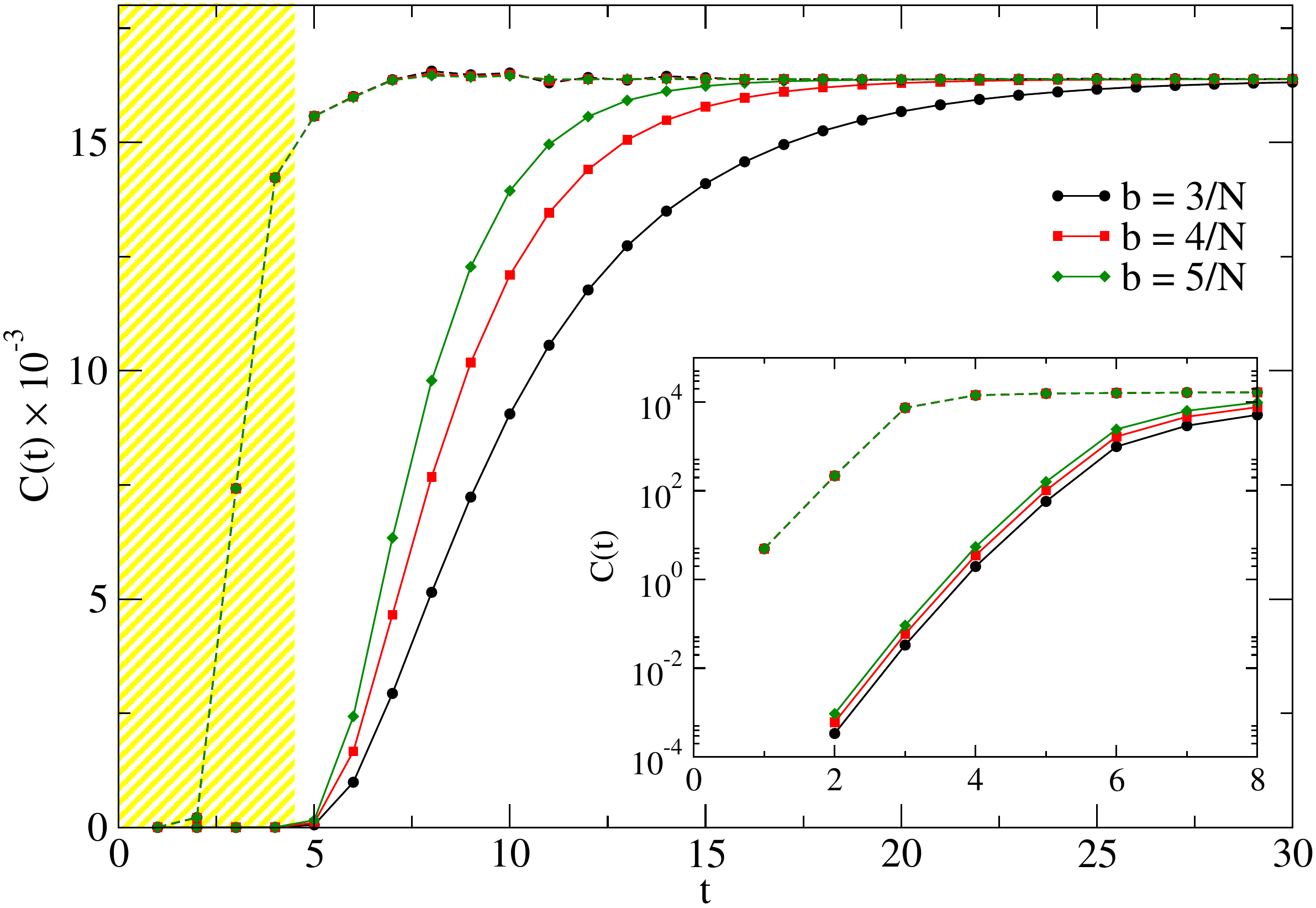}
\caption{$C(t)$ vs $t$ for coupled kicked rotor for strongly chaotic case with various values of interactions. The sub-system size $N$ is set to $256$. The interaction, $b$, scales as $1/N$. The kick parameters are $K_1 = 9$ and $K_2 = 10$. Inset corresponds to same plot in log-linear scale. The solid lines are for $C_{AB}(t)$, while the dashed ones are for $C_{AA}(t)$.
\label{fig-otoc-fft}}
\end{figure}

 The OTOCs $C_{AA}(t)$ and $C_{AB}(t)$ with $A(0)=\mathcal O_1 \otimes \mathbb I_2$ and $B(0)=\mathbb I_1 \otimes \mathcal O_2$ and $U$ the coupled standard map from Eq.~(\ref{eq-floquet}), are plotted in Fig.~(\ref{fig:K00},\ref{fig-otoc-fft-integrable}) for the regular case and Fig.~(\ref{fig-otoc-fft}) for the chaotic case. We see in all figures, except the $K_i=0$ case, two time
 regimes, one during which the growth is not really visible except on log-scales and the other during which substantial growth occurs. The time scale
 separating these two regimes is precisely the Ehrenfest time. Up to the Ehrenfest time, $t_E$, the OTOC exhibits a power law growth for non-chaotic and weakly chaotic cases. As shown in Fig.~(\ref{fig-otoc-fft-integrable}), the linear behavior in log-log scale confirms the power law growth. The $C_{AA}(t)$ follows quite simply from the classical counterpart of OTOC $C_{AB}(t)$ is $C_{cl}(t)$ where,
\begin{subequations}
\begin{align}
\label{eq-otoc-poisson}
\left< C_{cl}(t) \right>  & \equiv  \frac{\hbar^2}{4} \left< \{\cos(2\pi q_1(t)), \cos(2\pi q_2(0)) \}^2 \right> \\
& \propto \left< \left( \frac{\partial q_1(t)}{p_2(0)} \right)^2 \right>,
\end{align}
\end{subequations}
where $\left<.\right>$ represents the average over phase space points.

Taking the integrable $K_i=0$ case, the simplest situation is also zero interaction, which implies that the system is just two uncoupled free rotors. As for free motion we have $[\hat{q}(t), \hat{q}(0)] \sim t$, we
expect and find a quadratic growth of OTOC when the operators are in the same subspace, as shown in Fig.~(\ref{fig:K00}), however when the operators are in different subspaces, small interactions $\sim 1/N$ give rise to surprisingly large power laws $\sim t^5$, and this persists for higher values of the coupling constant $b$. The anomalously large power law  persists when the individual rotors are near-integrable and the interaction is small, as shown in Fig.~(\ref{fig-otoc-fft-integrable}). With growing interaction there is a transition to chaos and an approximately exponential growth is obtained. 
The near integrable regime of $K_1=0.5$ and $K_2=0.7$ is qualitatively similar with a power law  $\sim t^{2.1}$ growth of $C_{AA}(t)$ for $b=2/N$
and somewhat surprisingly an exponential growth was not obtained even at  $b=1.7$ when the spacing distribution is already Wigner, indicating the possibility that the OTOC are more sensitive to small regular regions than measures such as the nearest neighbor spacing statistics.


\newcolumntype{g}{>{\columncolor{color2}}c}
\begin{table}[t]
\begin{ruledtabular}
\begin{tabular}  {g g g}
\toprule
\rowcolor{color1}
\multicolumn{3}{c}{$K_1 = 9, K_2 = 10$}  \\
$N = 64$ & $N=256$ & {Classical}	\\
 $t_{EF} \approx 3$ & $t_{EF} \approx 4$ &							\\
 \toprule
$2\lambda_L = 3.91\pm0.01$& $2\lambda_L = 4.00\pm 0.02$& $2 \lambda_{cl} = 3.91$ \\
\rowcolor{color1}
\multicolumn{3}{c}{$K_1 = 19, K_2 = 20$} \\
$N=64$ & $N=256$ & {Classical}\\
$t_{EF}\approx 2$ & $t_{EF} \approx 3$ &  \\
 $2\lambda_L = 4.98\pm0.05$ & $2\lambda_L = 5.33\pm 0.01$ & $2 \lambda_{cl} = 5.34$\\
\rowcolor{color1}
\multicolumn{3}{c}{$K_1 = 20, K_2 = 21$} \\
$N=64$ & $N=256$ & {Classical}\\
$t_{EF} \approx 2$ & $t_{EF} \approx 3$ &  \\
 $2\lambda_L = 5.030\pm0.06$ & $2\lambda_L = 5.41\pm O(10^{-4})$ & $2 \lambda_{cl} = 5.44$
\end{tabular}
\end{ruledtabular}
\caption{The comparison of the quantum and classical Lyapunov exponents $\lambda_L$ and $\lambda_{cl}$ for various combinations of $N$ and $K_i$. The exponent is obtained by taking average for $1/N \le b \le 5/N$. 
\label{table-lambda-s}}
\end{table}

In contrast the case when both subsystems are strongly chaotic and the coupling is weak, the OTOC $C_{AB}(t)$ clearly shows two regimes, one wherein there is an initial exponential growth and then a gradual saturation as shown in Fig.~(\ref{fig-otoc-fft}).
For strongly chaotic case, the OTOC shows an exponential growth till the Ehrenfest time as $C_{AB}(t<t_{EF}) \propto b^2 \exp(2\lambda_L t)$. The behavior is clearly observed from plot for OTOC in inset of Fig.~(\ref{fig-otoc-fft}) in log-linear scale for strongly chaotic case. The dynamics is mainly governed by the sub-system chaos and is not affected by the interaction. For a chaotic system, if classical Lyapunov exponent is $\lambda_{cl}$, then we get,
\begin{align}
\label{otoc-exp}
C_{cl}(t) \propto b^2 \exp(2\lambda_{cl} t)
\end{align}
We numerically compare the quantum Lyapunov exponent from the OTOC computation, $\lambda_L$ with the classical exponent $\lambda_{cl}$. The classical Lyapunov exponent is calculated through the Poisson bracket. The exponent for a few cases are shown in Table \ref{table-lambda-s}. It is seen that the agreement between $\lambda_{cl}$ and $\lambda_L$ generally gets better for larger $N$ and is quite good. 

The Ehrenfest time shown in table is estimated from the fact that a cell of size $\hbar$ will take time $t_{EF}$ to spread over all the phase space, i.e, $\hbar \exp(\lambda_{cl}t_{EF}) \approx 1$ or $t_{EF} = \left| \log(\hbar) \right|/\lambda_{cl} = \log(N)/\lambda_{cl}$. We note that the interactions we have used are classically negligible, as they scales as $b \sim 1/N$, but is quantum mechanically large for the case when the subsystems are chaotic.
It is known that there is a transition to global quantum chaos and RMT behavior in the case for $b \sim 1/N^2$ \cite{Arul-PRL-2016}. Thus while these are small interactions,
there is already global quantum chaos and for example the eigenvalue statistics will be that of RMT. 

We also comment that the OTOC $C_{AA}(t)$ when both operators are in the same subspace is very different and is shown with dashed lines in 
Fig.~(\ref{fig-otoc-fft}). While there is still an exponential initial growth of the OTOC with almost identical Lyapunov exponent, 
there is practically saturation at the Ehrenfest time. Indeed we find that this OTOC is almost identical to that obtained
with one a single kicked rotor and differ only in the saturation value.

\begin{figure}[ht]
\centering
\subfloat[Subfigure 1 list of figures text][$K_1=0,\, K_2=10, \, b=0$]{
\includegraphics[width=0.7\textwidth]{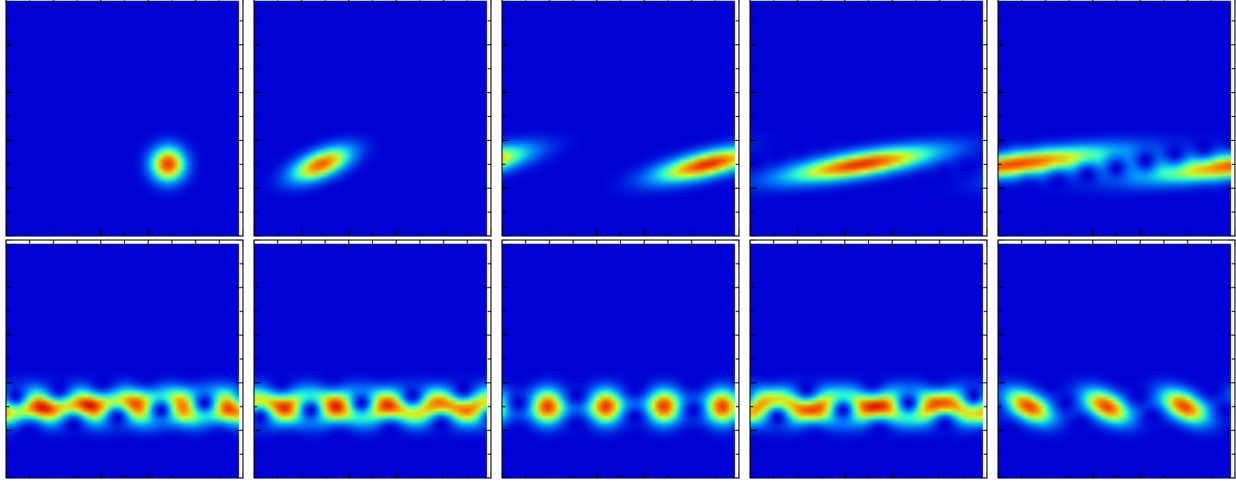}
\label{fig:subfig1}}
\qquad
\subfloat[Subfigure 1 list of figures text][$K_1=0,\, K_2=10, \, b=0.06$]{
\includegraphics[width=0.7\textwidth]{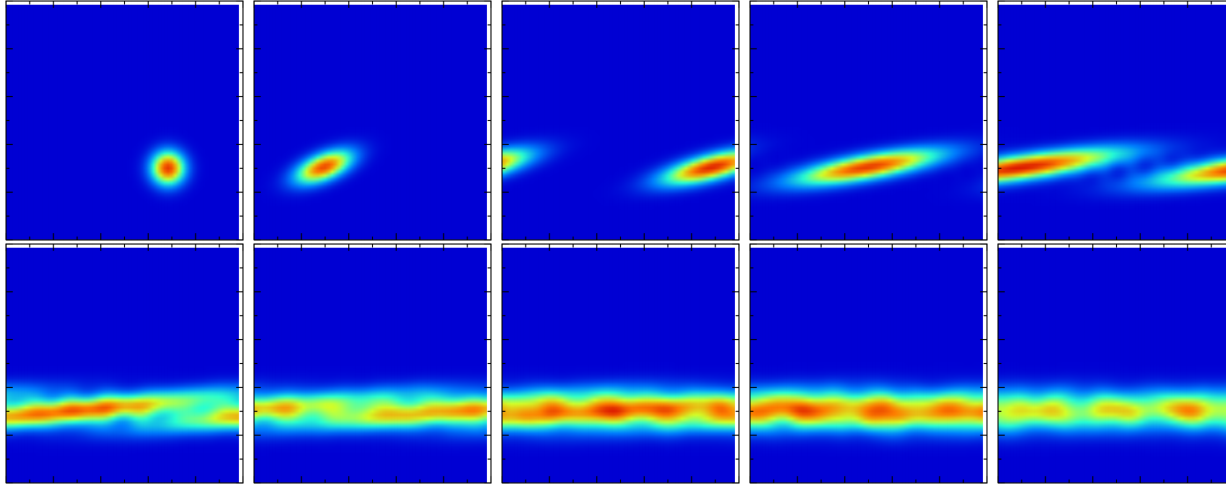}
\label{fig:subfig2}}
\qquad
\subfloat[Subfigure 1 list of figures text][$K_1=0,\, K_2=10, \, b=0.4$]{
\includegraphics[width=0.7\textwidth]{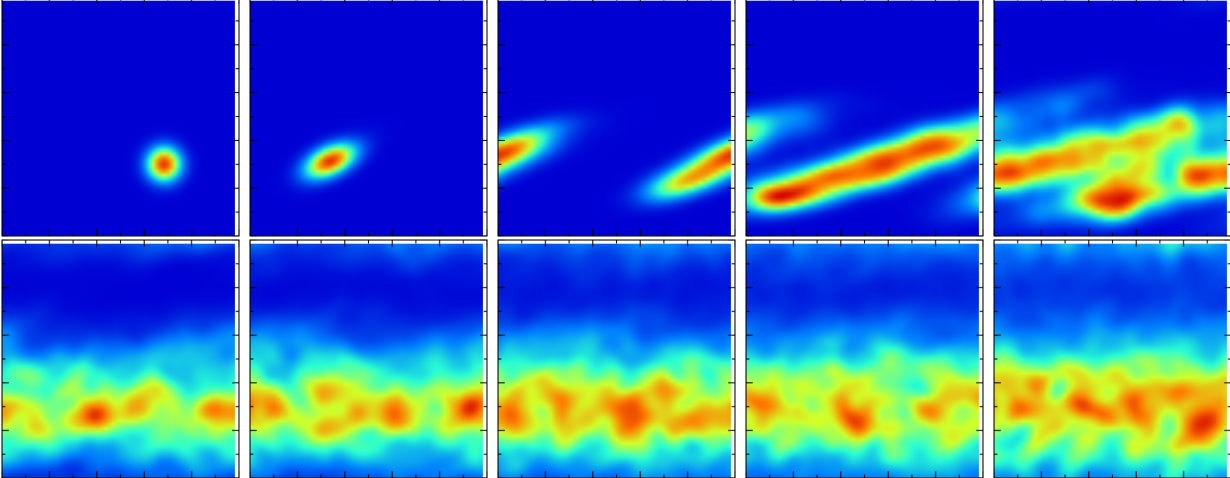}
\label{fig:subfig3}}
\caption{The Husimi, or coherent state, representation of the subsystem state in the $K_1=0$ rotor subspace. Shown are the states at times $0,2,4,6,8,12,14,16,18,20$, with $N=64$  in the unit $(q_1,p_1)$ square or torus. Notice the fractional revivals visible at zero interaction, getting smeared with increased interaction.}
\label{fig:HusimiK010}
\end{figure}

Finally we present preliminary results for the intriguing case $K_1=0$ and $K_2=10$, that is one of the subsystems is integrable
and the other is fully chaotic. For weak interactions, we expect the second subsystem to act as an agent of decoherence and destroy for example
quantum phenomena such as fractional revivals that occur in the $K_1=0$ subsystem. To visualize this we study the evolution of the $K_1=0$ subsystem Husimi functions, starting from a coherent state
localized at $(z_{10}, z_{20})=(q_{10},p_{10},q_{20},p_{20})$. We find 
$|\psi(n)\rangle=U^n |z_{10}\rangle |z_{20}\rangle$ and the subsystem state $\rho_1(n)=\text{Tr}_2(|\psi(n)\rangle \langle \psi(n)|)$ and its Husimi representation $\langle q_1 p_1|\rho_1(n)|q_1p_1 \rangle$, where $|z_1\rangle=|q_1p_1\rangle$ is a coherent state, a minimum uncertainty state centered at $z_1=(q_1,p_1)$. Roughly speaking it is the shadow of the state in subsystem $1$. This is shown in Fig.~(\ref{fig:HusimiK010}) for 3 values of the interaction $b$, on 
the phase-space unit square. When $b=0$ as in Fig.~(\ref{fig:subfig1}), the system is simply a free particle on a ring, and there is the phenomena of fractional revivals when the density forms itself into several spatially localized ``cat states" like patterns. There is an initial time, the Ehrenfest time, which scales as $\sqrt{N}$ before which the quantum interference effects are negligible. When the interactions are turned on the behaviour in the pre-Ehrenfest time is not changed, but in the post-Ehrenfest time, the coherent interference effects giving rise to revivals gets destroyed.

\begin{figure}[!h]
\centering
\subfloat[Subfigure 1 list of figures text][The OTOC when observables are in the same (regular, $K_1=0$) subspace $C_{AA}(t)$.]{
\includegraphics[width=0.32\textwidth]{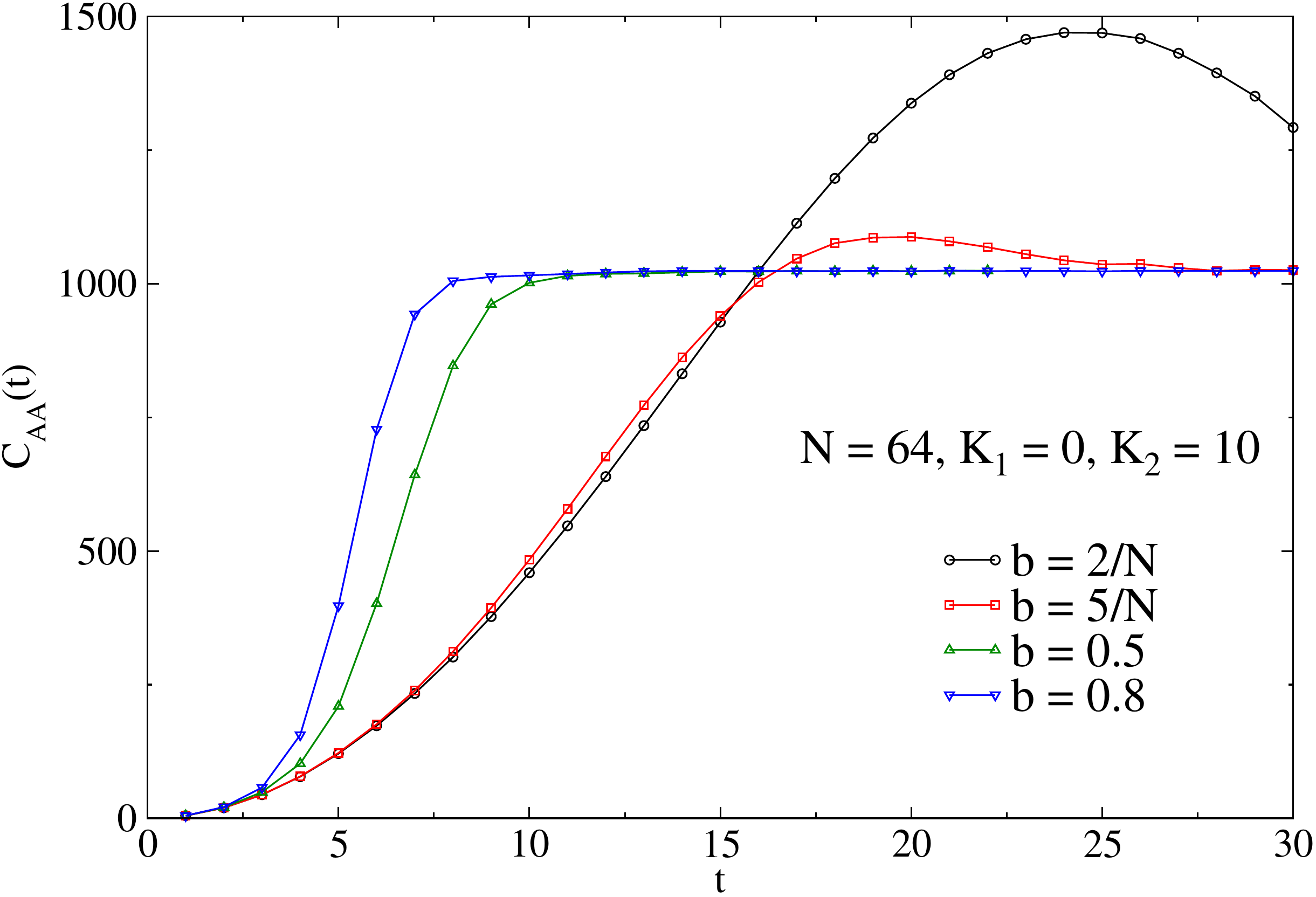}
\includegraphics[width=0.32\textwidth]{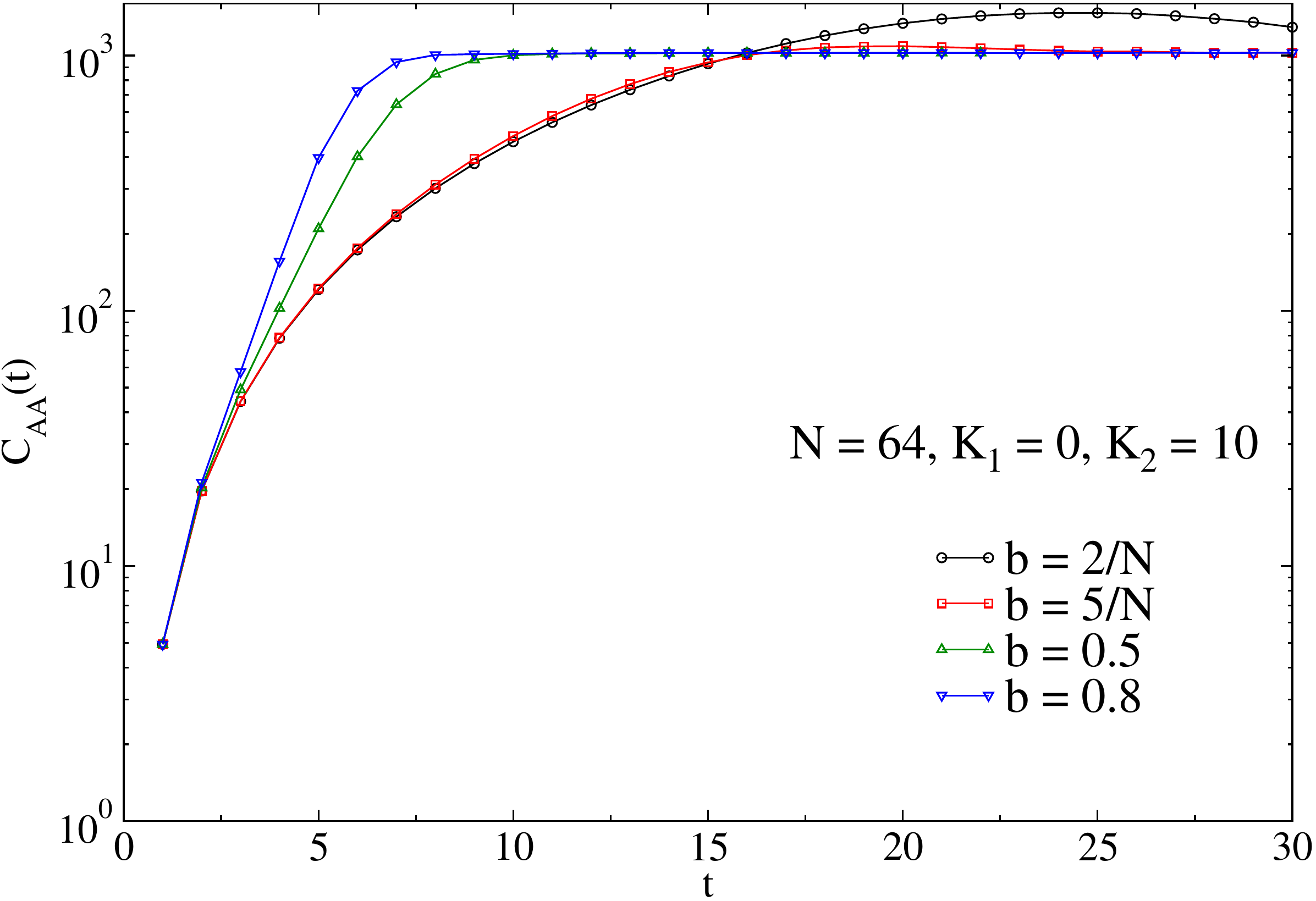}
\includegraphics[width=0.32\textwidth]{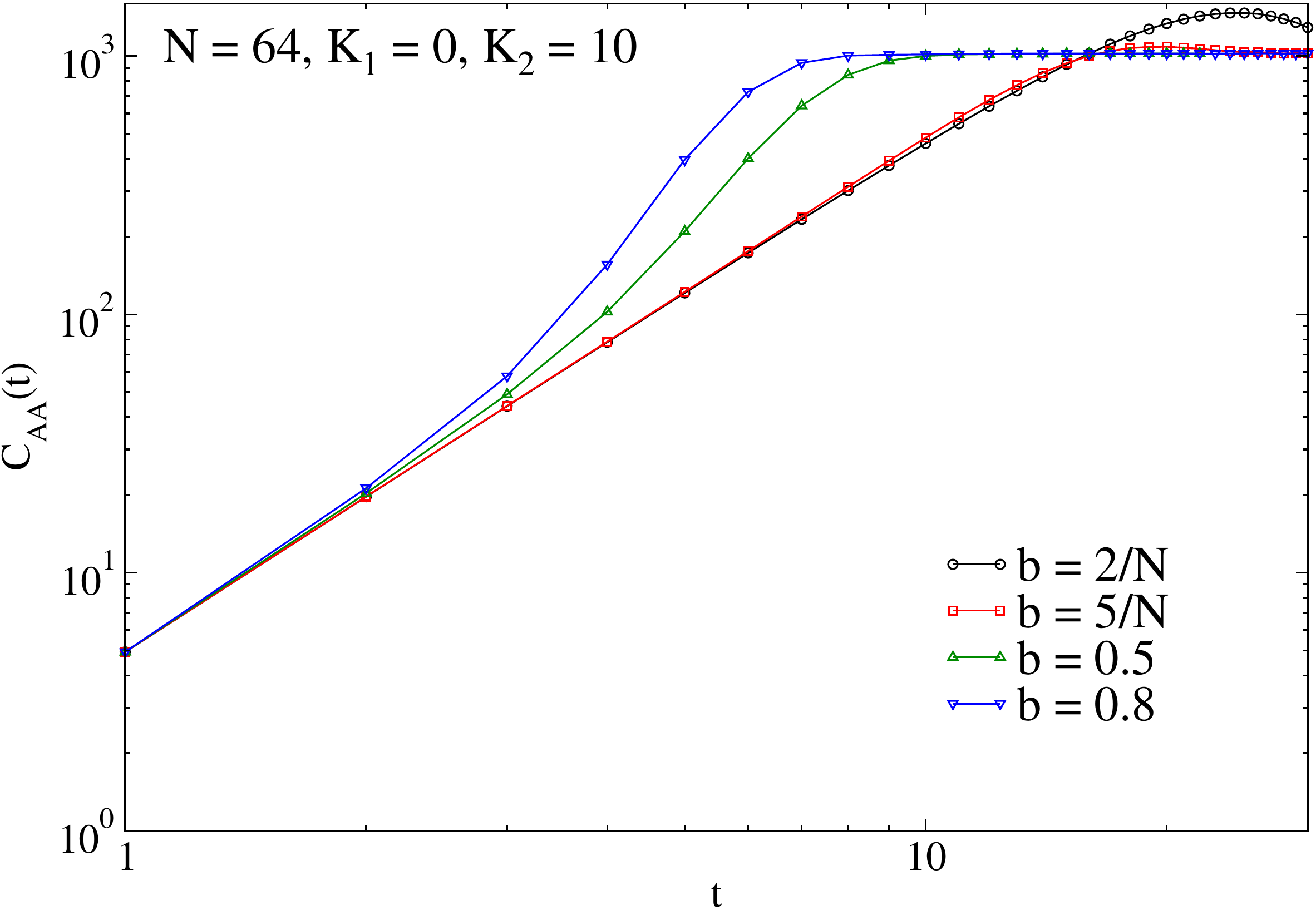}
\label{fig:K0K10a}}
\qquad
\subfloat[Subfigure 1 list of figures text][The OTOC when observables are in different subspaces $C_{AB}(t)$.]{
\includegraphics[width=0.32\textwidth]{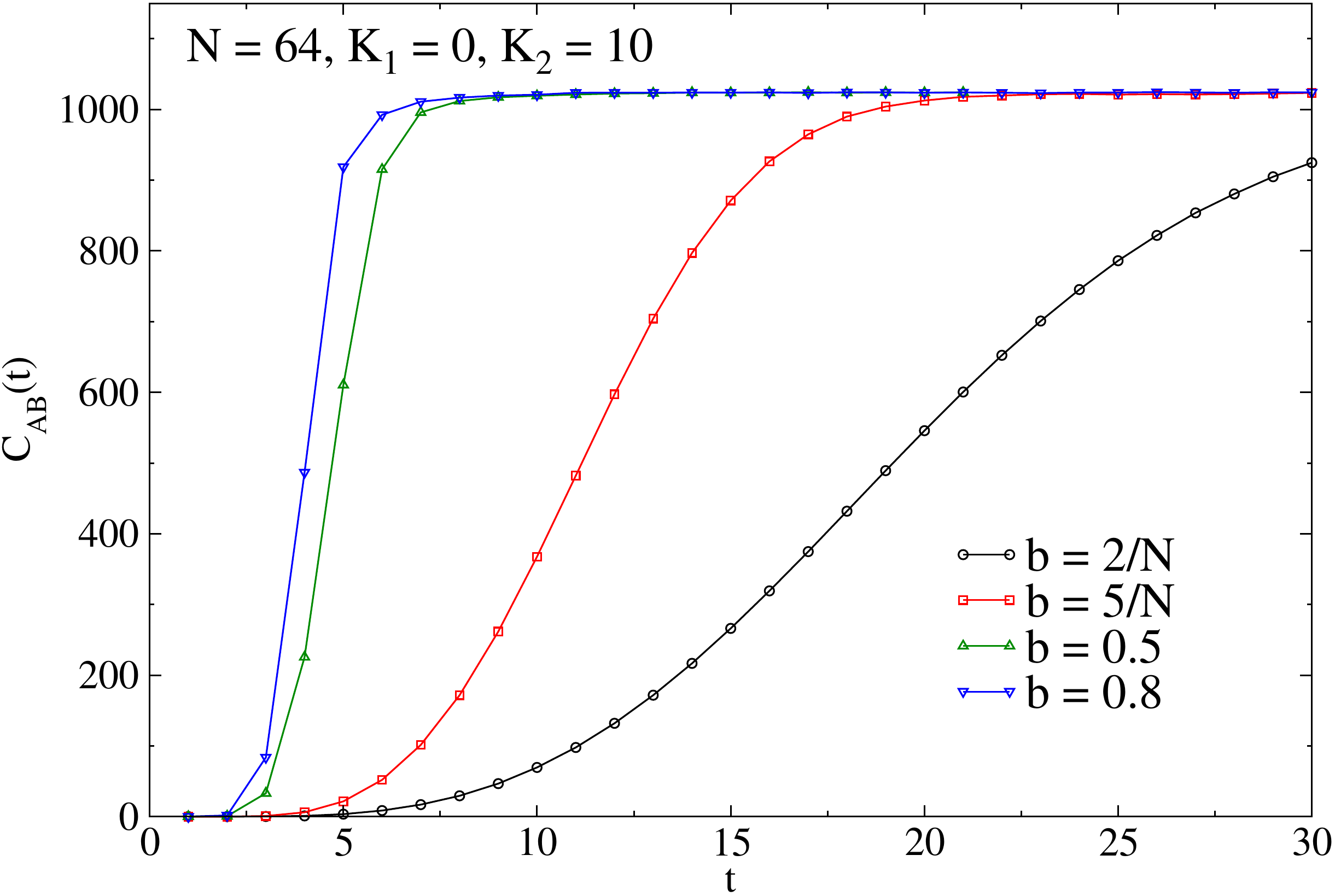}
\includegraphics[width=0.32\textwidth]{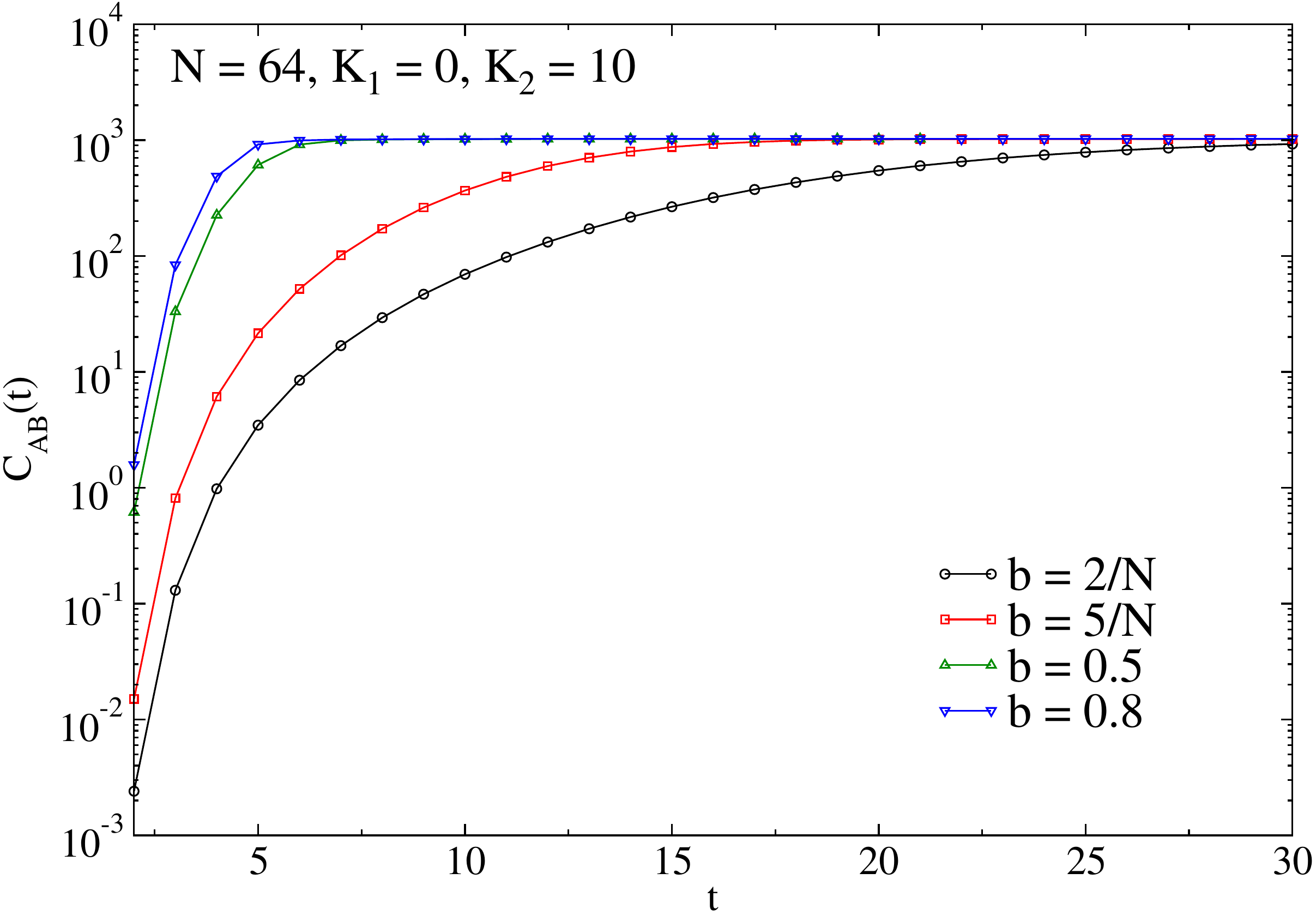}
\includegraphics[width=0.32\textwidth]{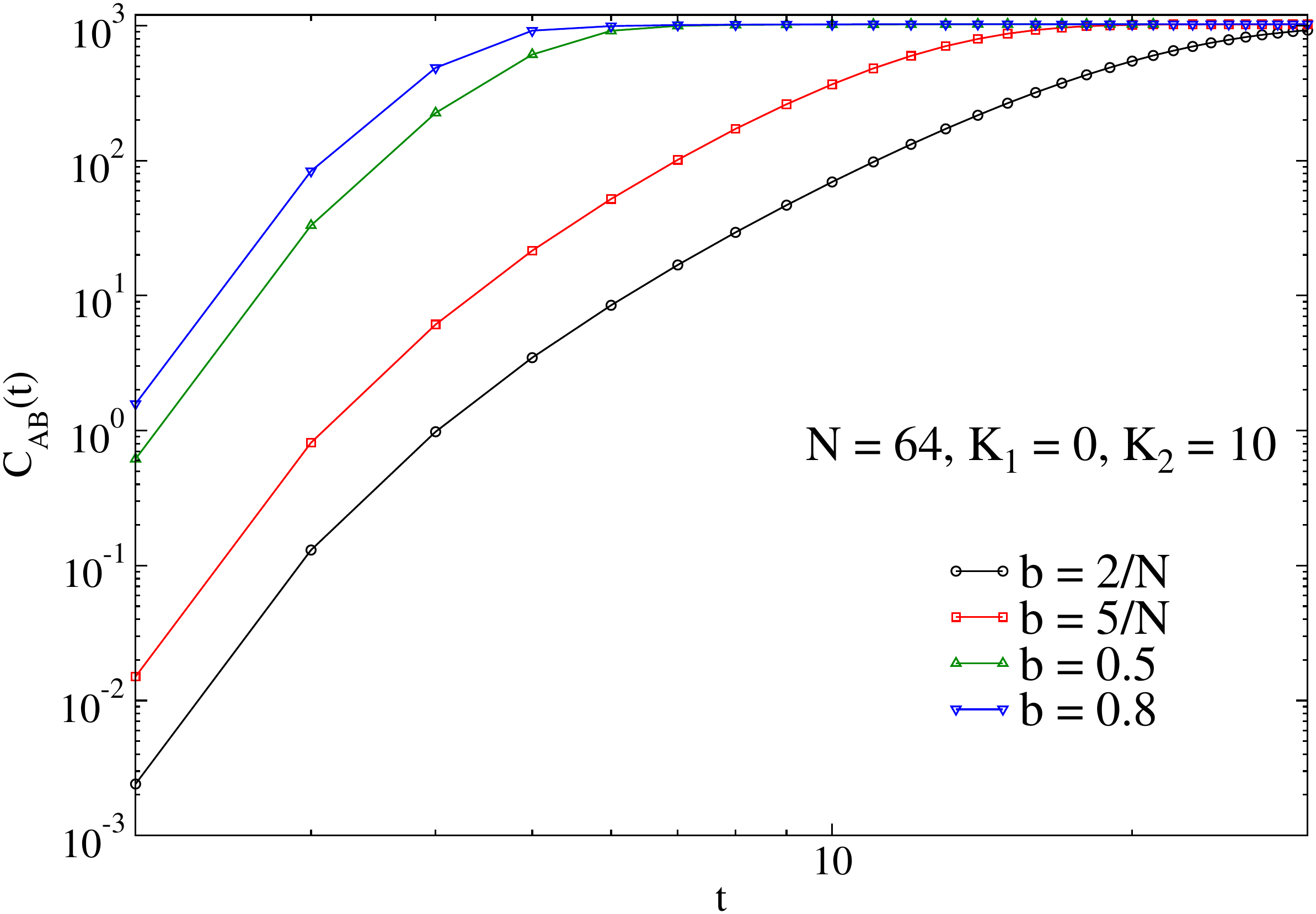}
\label{fig:K0K10b}}
\caption{The OTOCs for the case when $K_1=0$ and $K_2=10$.
Shown in each case are the same data in linear-linear, linear-log, and log-log plots.
}
\label{fig:OTOCK0K10}
\end{figure}

These have their imprints on the OTOCs of observables as seen in Fig.~(\ref{fig:OTOCK0K10}). We notice from Fig.~(\ref{fig:K0K10a}) that if the interactions are small $b \sim 1/N$, the OTOCs $C_{AA}$, when the observables are in the ``regular" subspace are approximately growing as $t^2$ and are practically independent of the interaction, and do not seem to show differences at the Ehrenfest time. With increasing interaction the OTOC grow much faster and saturate at the Ehrenfest time, marking the onset of decoherence. Not shown is the case when both the operators are in the chaotic subsystem, when the growth is exponential. Thus the interesting case of $C_{AB}(t)$ is shown in Fig.~(\ref{fig:K0K10b}), where we now see that even small interactions can be distinguished due to the dependence on $b^2$ as before. However for small interactions, the growth is neither a clear exponential nor a power-law one. With increased interaction, the OTOC is smaller than the chaotic-chaotic case but the growth rate is nearly exponential. The saturation time also decreases with increasing interaction indicating the decrease in the Ehrenfest time. We notice that these preliminary studies throw up several interesting questions, including the very definition of Ehrenfest time scales in such multipartite systems.

\section{An RMT model and the post-Ehrenfest regime in the chaotic-chaotic case} \label{sec-rmt-model}
While the pre-Ehrenfest regime is interesting in many cases, especially when there is a mixture of regular and chaotic subsystems, the post-Ehrenfest time is less understood.
See \cite{Klaus-2018,Saraceno-2018} for recent works concerning this. In this section we summarize a work \cite{Ravi2019} that provides a complete theory for the case when both subsystems are chaotic. 
As seen in Fig.~(\ref{fig-otoc-fft}), the pre-Ehrenfest time is marked by an exponential 
growth which is essentially coming from classical Poisson brackets. However, this correspondence breaks down if the operators themselves do not have a smooth classical limit. It was seen that the observables which have no classical counterpart skip the Lyapunov regime and start relaxing exponentially to the saturation value. 
The post-Ehrenfest regime of smooth operators is also exactly of the same kind and is universal in the sense that the rate does not depend on the characteristics of sub-systems. This is reasonable as a smooth operator has been scrambled sufficiently by the 
Ehrenfest time to resemble generic operators.

While the relaxation to saturation is via a power-law for integrable and weakly-chaotic systems it is exponential for fully chaotic systems. 
In the last case, we determine the relaxation rate by replacing the subsystem dynamics with random unitary matrices that are independent at each time step. This leads to the OTOC estimate
\begin{align}
C_{AB}(t > t_{EF}) = C_\infty\left[1 - \gamma(b) e^{-\mu(b)(t-t_{EF})}\right]
\end{align}
where $\mu(b)$ is the relaxation rate that depends on the exact nature of the interaction in the Hamiltonian. For coupled kicked rotors the relaxation rate is given by,
\begin{align}
\label{eq-mu-qkr}
\mu(b) = \ln \left | J_0\left(\frac{Nb}{2\pi}\right)\right|^{-4} \approx \frac{N^2 b^2}{4 \pi^2}, 
\end{align}
where $J_0(x)$ is the Bessel function of the first kind.
\begin{figure}[!h]
\centering
\includegraphics[width=0.45\textwidth]{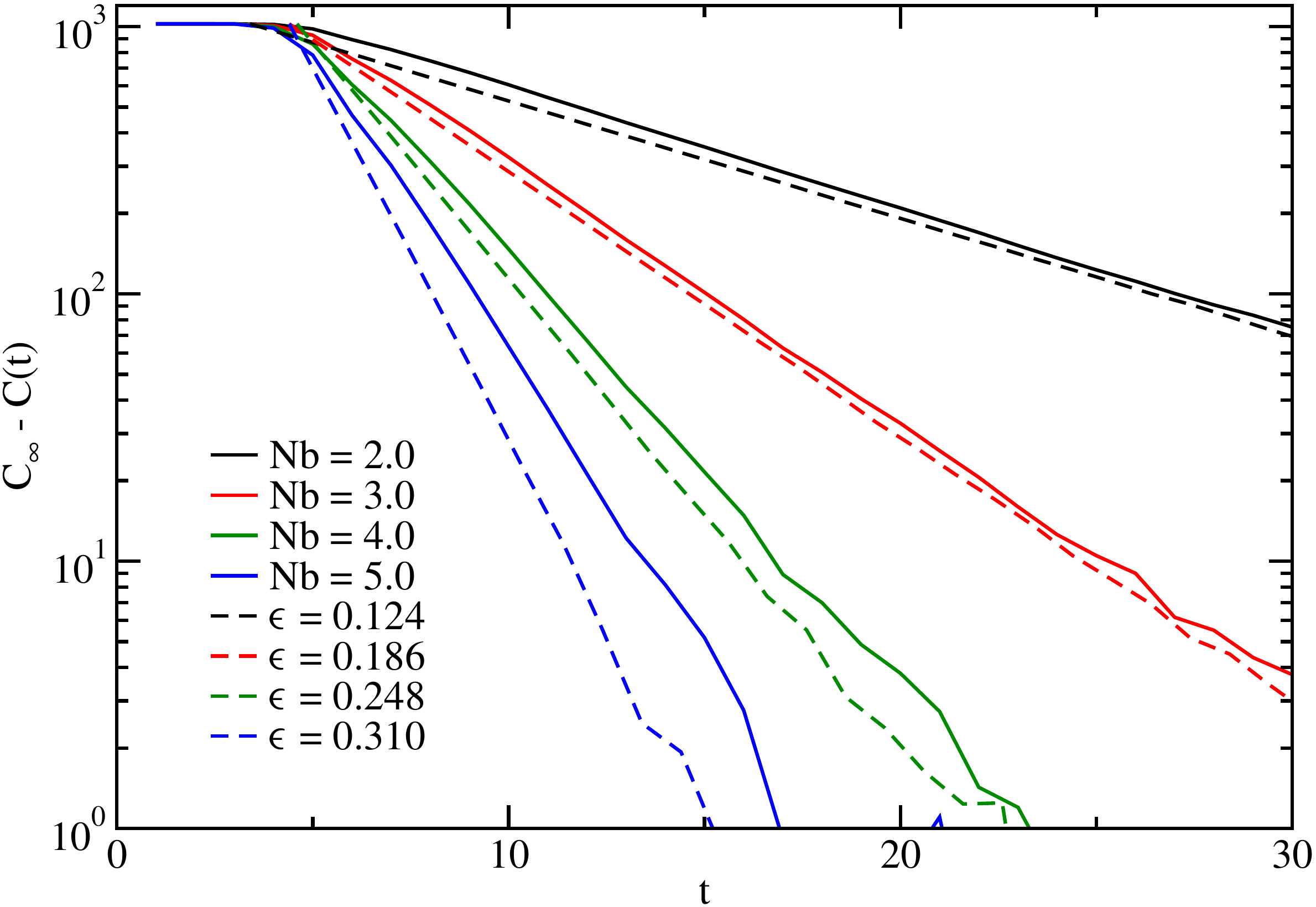}
	\includegraphics[width=0.45\textwidth]{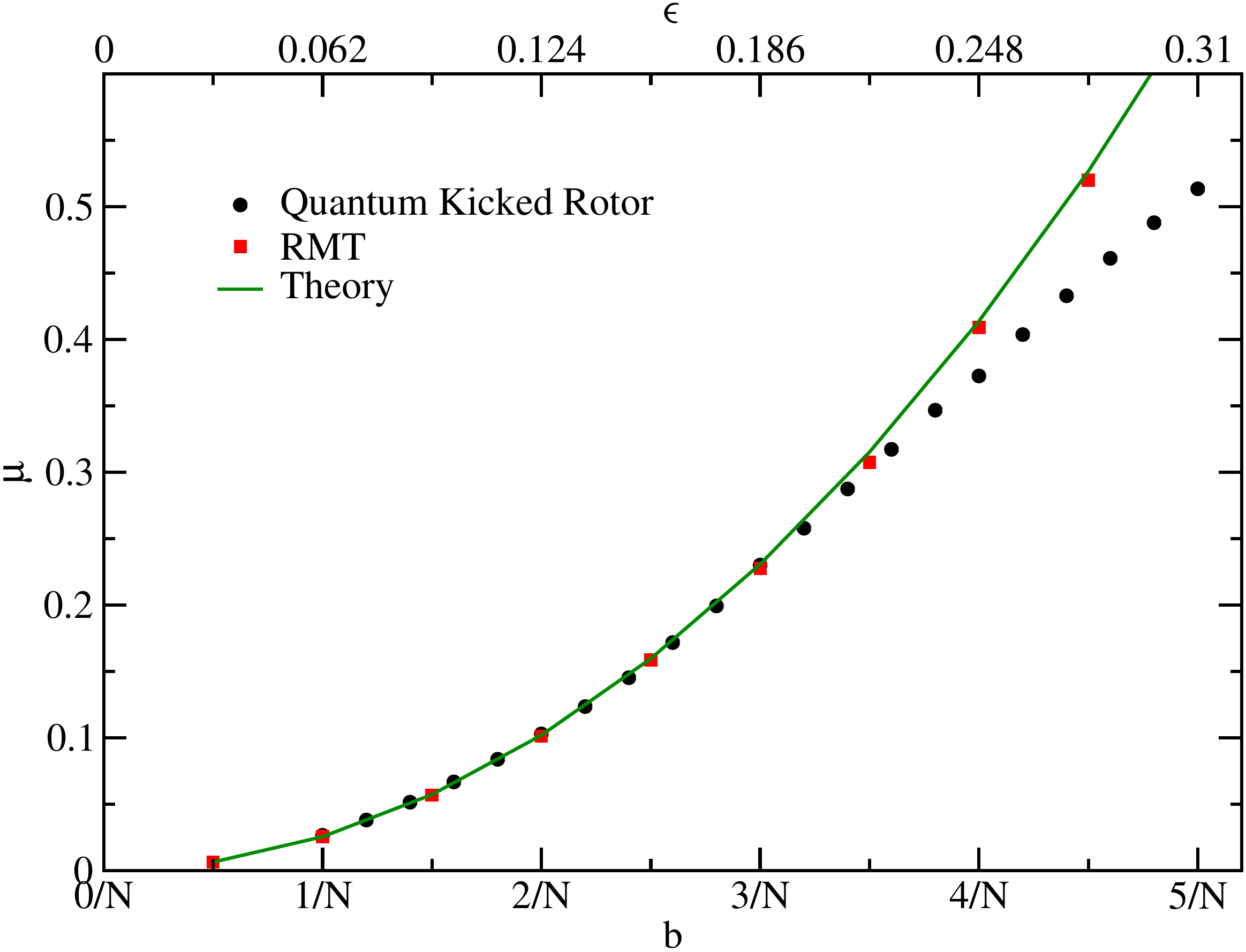}
\caption{$\log(C_{\infty}-C_{AB}(t))$ v.s. $t$ for the coupled kicked rotor and corresponding RMT model. The subsystem size, $N = 64$ is considered here. Note that the RMT results (dashed lines) are shifted arbitrarily for an easy comparison with the coupled kicked rotor's results. The rate $\mu$ \emph{vs} interaction $Nb$ and $\epsilon$.\label{fig-post-ln-otoc-fft-rmt}}
\end{figure}
This implies that $\ln(C_\infty - C(t)) \propto -\mu(b)t$ and we show in Fig.~(\ref{fig-post-ln-otoc-fft-rmt}) the relaxation rate of quantum kicked rotor and its comparison with Eq.~(\ref{eq-mu-qkr}). We observe an excellent qualitative agreement, after the Ehrenfest time.

To model the evolution due to the propagator in Eq.~(\ref{eq-floquet}), as the subsystem dynamics is chaotic, the operators $\mathcal F_j$ maybe considered to be chosen from the circular ensemble of random matrices, that is
\begin{align}
\label{eq-floquet-rmt}
\mathcal F_1, \mathcal F_2 \equiv \text{COE or CUE},
\end{align}
where the circular orthogonal/unitary ensemble (COE/CUE) applies when there is/is no time-reversal symmetry. The interaction $U_b$ is diagonal in the case of the standard maps in the position representation. We therefore model a random interaction as a diagonal unitary random matrix, which is just a diagonal matrix of pure phases. We take these $N^2$ phases to be of the form $\exp( i \epsilon \xi)$ where $\xi \in [-\pi,\pi)$ is 
uniformly random. The strength of the interaction is determined by $\epsilon$. If $\epsilon=0$, this is noninteracting and the case $\epsilon=1$ is one of maximal interaction. Calling such a diagonal matrix $U_\epsilon$, we model the matrix power $\mathcal U^t$ by 
\begin{equation}
\mathcal U^{(t)} = \prod_{j=1}^t (\mathcal{F}_{1j}\otimes \mathcal{F}_{2j}) \mathcal U_{\epsilon j},
\end{equation}
where the index $j$ implies that at each step we choose different locals
as well as interaction matrices $\mathcal U_{\epsilon}$. The entire physics then rests on the strength of the interaction $\epsilon$ and the observables. We expect such a model to work well if the interaction is not so weak that it does not mix unperturbed levels at all, which in the case of kicked rotors would be $b \gg 1/N^2$.

To derive an expression for the OTOC $C_{AB}(t)$, we consider instead of $\mathcal U^t$ the quantity $\mathcal U^{(t)}$. We then average over the $\mathcal F_{1j}$, $\mathcal F_{2j}$ to get \cite{Ravi2019} the two-point correlator:
\begin{align}
C_2(t) \approx \overline{C_2(t)}=	 C_\infty =\Tr(\mathcal O_1^2) \Tr(\mathcal O_1^2)
\end{align}
and the 4-point out-of time-ordered correlator
\begin{align}
\label{eq-c4-expression}
C_4(t) \approx \overline{C_4(t)}=\sinc^{4t}(\pi \epsilon) ~ \Tr (\mathcal O_1^2) ~\Tr (\mathcal O_2^2) ,
\end{align}
which leads to the RMT model OTOC:
\begin{equation}
\label{eq-otoc-formula}
C_{AB}(t) = \Tr(\mathcal O_1^2)~ \Tr(\mathcal O_2^2) \left[ 1- \sinc^{4(t-1)}(\pi \epsilon)\right], \; t\geq 1.
\end{equation}
Here we have taken into account a detail that is essential for the observables $\mathcal{O}$ which we take to be diagonal in the same basis
as the interaction is. The OTOC of the random matrix model therefore approaches saturation $C_{\infty}$ exponentially with a rate $\mu_{RMT}(\epsilon)=-4 \ln |\sinc(\pi \epsilon)| \approx 2 \pi^2 \epsilon^2/3$ that is universal in the sense that it is independent of the choice of operators and depends only on the interaction. We see clearly that the two-point part of the OTOC $C_2(t)$ contains essentially no interesting behaviour being approximately a constant, while the 4-point correlator $C_4(t)$ contains all the non-trivial information.
The random matrix model can be used for non-random interactions with diagonal matrix elements of the form $\exp(-i \epsilon V_{mn})$, where
$V$ is the interaction potential or Hamiltonian. This leads to the rate
\begin{align}
\mu(\epsilon) = -4 \ln \left | \int_0^1 d\xi_1 d\xi_2 e^{-i \epsilon V(\xi_1,\xi_2)t/h}\right |,
\end{align}
and for the coupled standard map with $V = V_b = \cos(2\pi (q_1 + q_2))/4\pi^2$ the relaxation rate is same as in Eq.~(\ref{eq-mu-qkr}). Figure~(\ref{fig-post-ln-otoc-fft-rmt}) illustrates both the coupled standard map and the RMT models and also contains a comparison of the 
numerically obtained rates with the estimate $\mu(\epsilon)$. It is seen that the estimate breaks down for large coupling and a more complete theory is needed to account for the rate in these cases where the RMT 
does an overestimate.

\section{Summary and Outlook}\label{sec-conclusion}
We have studied OTOCs in a bipartite system of two coupled standard maps.
It is a rich model that allows for us to study the case when the subsystems are both chaotic, or when one is chaotic and the other regular and when both are regular. We have presented essentially numerical and preliminary results for two kinds of OTOCs in these systems. While power laws understandably dominate the growth of the OTOC in regular-regular systems, the case of regular-chaotic systems is unclear. This is an interesting case as it could model open or noisy regular systems.
The case of chaotic-chaotic systems are described by random matrix theory and appropriate models allow us to derive a universal exponential decay in
the post-Ehrenfest regime. The universality is in the form of the decay and the rates being independent of the observables used in the OTOC. Ehrenfest time itself needs to be more carefully studied in the case of weakly coupled systems as they could vary 
dramatically between the subsystems themselves.  
While we have restricted ourselves to the simplest many-body situation, naturally extensions to tripartite and multipartite systems is 
interesting and we hope that our study will add in some way to the vast literature that is emerging on the many facets of the OTOC.

\begin{acknowledgments}
RP acknowledges the SERB NPDF scheme (File No. PDF/2016/002900) for financial support. AL is grateful to Karol {\.Z}yczkowski for
the visits to Krakow and Warsaw during the Summer of 2019, and to the organizers of the Workshop on Quantum Chaos and Localization phenomena for the opportunity to attend the stimulating meeting and present parts of this work.
\end{acknowledgments}

\bibliography{references.bib}

\end{document}